\documentclass[onecolumn,showpacs,preprintnumbers]{revtex4}
\usepackage{mathrsfs}
\usepackage{graphicx}
\usepackage{dcolumn}
\usepackage{bm}
\usepackage{epsfig}
\usepackage{amsmath,amssymb}

\begin{document}
\title{Analysis of the strong vertexes of $\Sigma_c^{*} ND$ and $\Sigma_b^{*}NB$ in QCD sum rules}
\author{Guo-Liang Yu$^{1}$}
\email{yuguoliang2011@163.com}
\author{Zhi-Gang Wang$^{1}$}
\email{zgwang@aliyun.com}
\author{Zhen-Yu Li$^{2}$}

\affiliation{$^1$ Department of Mathematics and Physics, North China
Electric power university, Baoding 071003, People's Republic of
China \\$^2$ School of Physics and Electronic Science, Guizhou
Normal College, Guiyang 550018, People's Republic of China}
\date{\today }

\begin{abstract}
In this article, we analyze the strong vertexes $\Sigma_{c}^{*}ND$
and $\Sigma_{b}^{*}NB$ using the three-point QCD sum rules under the
Dirac structure of $q\!\!\!/p\!\!\!/\gamma_{\mu}$. We perform our
analysis by considering the contributions of the perturbative part
and the condensate terms of $<\overline{q}q>$ and
$<\frac{\alpha_{s}}{\pi}GG>$. After the form factors are calculated,
they are then fitted into analytical functions which are used to get
the strong coupling constants for these two vertexes. The final
results are $g_{\Sigma_{c}^{*}ND}=7.19^{+8.49}_{-3.11}\pm1.76$ and
$g_{\Sigma_{b}^{*}NB}=10.54^{+15.59}_{-5.23}\pm1.82$.
\end{abstract}


\maketitle

\begin{large}
\textbf{1 Introduction}
\end{large}

The charmed and bottom baryons, which contain at least a heavy
quark, serve as a particular laboratory for studying dynamics of the
light quarks in the presence of the heavy quark(s), and also as an
excellent ground for testing predictions of the quark model and
heavy quark symmetry. The properties of these heavy baryon states
mainly include the mass spectrum, the magnetic moments, the strong,
electromagnetic and weak decay behaviors. Investigation of these
properties can give us useful information on the quark structure of
these baryons. The strong coupling constants is an important
parameter about the strong interactions of the heavy baryons. The
accurate determination of the coupling constants can not only help
us further understanding the strong decay behaviors of these heavy
baryons but also give us the knowledge about its nature and
structure.

By this time, many heavy baryon states have been discovered in
experiments by BaBar, Belle, CDF and D0
Collaborations\cite{Aube06,Naka10,Lesi,Rosn07}. These states include
the charmed baryons such as $\frac{1}{2}^{+}$ antitriplet
states($\Lambda^{+}_{c}$,$\Xi^{+}_{c}$,$\Xi^{0}_{c}$), the
$\frac{1}{2}^{+}$ and $\frac{3}{2}^{+}$ sextet states
($\Omega_{c}$,$\Sigma_{c}$,$\Sigma'_{c}$) and
($\Omega^{*}_{c}$,$\Sigma^{*}_{c}$,$\Sigma^{*}_{c}$)\cite{Naka10}.
Besides, CDF and LHCb Collaborations observed several S-wave bottom
baryon states, $\Lambda_{b}$,$\Sigma_{b}$,$\Sigma^{*}_{b}$,$\Xi_{b}$
and $\Omega_{b}$\cite{Paul,Klem10}. The SELEX collaboration even
reported the observation of the signal for the doubly charmed baryon
state $\Xi_{cc}^{+}$\cite{Matt,Oche}. Stimulated by these
discoveries, theorists studied the nature of these baryons with
different theoretical
approaches\cite{Faes,Pate,LiuX,ChenHX1,Chun,ChenW,Zhjr,Karl1,Karl2,Nava,Khod,Alie,Azi1,Azi2,Wzg,KangXW,Esposito}.
As mentioned above, the subsequent analysis of the strong decays of
these baryons requires knowledge about their strong coupling
constants. Thus, people calculated some of the strong coupling
constants,
$g_{\Omega^{*}_{c}\Omega^{*}_{c}\phi}$,$g_{\Omega^{*}_{c}\Xi^{*}_{c}K^{*}}$,
$g_{\Xi^{*}_{c}\Sigma^{*}_{c}K^{*}}$,$g_{\Omega^{*}_{b}\Omega^{*}_{b}\phi}$,$g_{\Xi^{*}_{b}\Sigma^{*}_{b}K^{*}}$,$g_{\Xi^{*}_{b}\Xi_{b}\pi}$,$g_{\Xi^{*}_{c}\Xi_{c}\pi}$,
$g_{\Lambda_{b}NB^{*}}$,$g_{\Lambda_{b}N^{*}B^{*}}$,$g_{\Lambda_{c}ND^{*}}$,$g_{\Lambda_{c}N^{*}D^{*}}$,$g_{\Sigma_{c}ND^{*}}$,$g_{\Sigma_{b}N^{*}B^{*}}$,
$g_{\Sigma_{b}NB^{*}}$ and $g_{\Sigma_{c}N^{*}D^{*}}$,
etc\cite{Azi1,Azi2,Wzg5,Navar,Khodj,GuoLY2,Azizi4}. For these work,
QCD sum rules proved to be a most powerful nonperturbative method
which has been widely used to analyze the properties of the
hadrons\cite{Brac1,Brac2,Alie2,Alie3,Alie4,Doi,Altm,Wzg3,Cerq,Rodr,Yazi,Khos1,Khos2,Rein,Pasc,Wzg4,Guoly}.

In the present paper, we calculate the strong coupling constants
$g_{\Sigma_{c}^{*}ND}$ and $g_{\Sigma_{b}^{*}NB}$ within the
framework of the QCD sum rules. The results of this work are
relevant in the bottom and charmed meson cloud description of the
nucleon which may be used to explain exotic events observed by
different collaborations. Besides, the exact values of these strong
coupling constants are essential to determine the modifications on
the masses, decay constants and other parameters of the $B$ and $D$
mesons in nuclear medium\cite{WzgH,Kuma,Haya}. The layout of this
paper is as follows. The next section presents the details of the
analysis of the vertexes $\Sigma_{c}^{*}ND$ and
 $\Sigma_{b}^{*}NB$. In Sec.3, we
present the numerical results and discussions, and Sec.4 is reserved
for our conclusions.

\begin{large}
\textbf{2 QCD sum rules for $\Sigma_{c}^{*}ND$ and
 $\Sigma_{b}^{*}NB$}
\end{large}

We study the strong coupling constants of the vertexes
$\Sigma_{c}^{*}ND$ and $\Sigma_{b}^{*}NB$ with the following
three-point correlation function,
\begin{eqnarray}
\Pi_{\mu}(p,p',q)=i^{2}\int d^{4}x\int
d^{4}ye^{-ip.x}e^{ip'.y}\Big\langle0|\mathcal
{T}\Big(J_{N}(y)J_{D[B]}(0)\overline{J}_{\mu\Sigma_{c}^{*}[\Sigma_{b}^{*}]}(x)\Big)
|0\Big\rangle,
\end{eqnarray}
where $\mathcal {T}$ is the time ordered product and
$J_{\mu\Sigma_{c}^{*}[\Sigma_{b}^{*}]}$, $J_{N}$ and $J_{D[B]}$ are
the interpolating currents of the baryons
$\Sigma_{c}^{*}[\Sigma_{b}^{*}]$, $N$ and the meson $D[B]$. Baryon
current is a composite operator with the same quantum numbers as a
given baryon, which include several possibilities\cite{Ioff1,Cola}.
For simplicity, the interpolating currents used in Equation(1) are
written as the following form,
\begin{eqnarray}
\notag
&& J_{\mu\Sigma_{c}^{*}[\sum_{b}^{*}]}(x)=\epsilon_{ijk}\Big(u^{iT}(x)C\gamma_{\mu}d^{j}(x)\Big)c[b]^{k}(x), \\
\notag
&& J_{N}(y)=\epsilon_{ijk}\Big(u^{iT}(y)C\gamma_{\mu}u^{j}(y)\Big)\gamma_{5}\gamma^{\mu}d^{k}(y),\\
&& J_{D[B]}(0)=\overline{u}(0)\gamma_{5}c[b](0).
\end{eqnarray}
The correlation function will be calculated in two different ways,
from the hadron degrees of freedom and quark degrees of freedom,
which are called the phenomenological side and the operator product
expansion(OPE) side separately.

\begin{large}
\textbf{2.1 The phenomenological side}
\end{large}

In order to obtain the phenomenological representations, we insert a
complete set of intermediate hadronic states with the same quantum
numbers as the current operators
$J_{\mu\Sigma_{c}^{*}[\Sigma_{b}^{*}]}$, $J_{N}$ and $J_{D[B]}$ into
the correlation $\Pi_{\mu}(p,p',q)$. Then, after the ground-state
contributions are isolated, we get the following function.
\begin{eqnarray}
 \notag
 \Pi^{HAD}_{\mu}(p,p',q)=&&\frac{\Big \langle 0| J_{N}|N(p')\Big
\rangle \Big \langle 0| J_{D[B]}|D[B](q)\Big \rangle \Big \langle
\Sigma_{c}^{*}[\Sigma_{b}^{*}](p)|
\overline{J}_{\mu\Sigma_{c}^{*}[\Sigma_{b}^{*}]}|0\Big
\rangle}{(p^{2}-m^{2}_{\Sigma_{c}^{*}[\Sigma_{b}^{*}]})(p'^{2}-m^{2}_{N})(q^{2}-m^{2}_{D[B]})}\\
&& \Big\langle
N(p')D[B](q)|\Sigma_{c}^{*}[\Sigma_{b}^{*}](p)\Big\rangle+\cdots
\end{eqnarray}
Where $h.r.$ stands for the contributions of the higher resonances
and continuum states. After the matrix elements appearing in the
above equation are substituted for the following parameterized
equations,
\begin{eqnarray}
\notag
&&\langle 0| J_{N}|N(p') \rangle=\lambda_{N}u_{N}(p',s'),\\
\notag
&&\langle0|J_{D[B]}|D[B](q) \rangle=i\frac{m_{D[B]}^{2}f_{D[B]}}{m_{u}+m_{c[b]}},\\
\notag &&\langle
\Sigma_{c}^{*}[\Sigma_{b}^{*}](p)|\overline{J}_{\mu\Sigma_{c}^{*}[\Sigma_{b}^{*}]}|0
\rangle=\lambda_{\Sigma_{c}^{*}[\Sigma_{b}^{*}]}\overline{u}_{\mu\Sigma_{c}^{*}[\Sigma_{b}^{*}]}(p,s),\\
&&\langle
N(p')D[B](q)|\Sigma_{c}^{*}[\Sigma_{b}^{*}](p)\rangle=ig_{\Sigma_{c}^{*}ND[\Sigma_{b}^{*}NB]}\overline{u}_{N}(p',s')u_{\Sigma_{c}^{*}[\Sigma_{b}^{*}]\alpha}(p,s)q^{\alpha},
\end{eqnarray}
the correlation function  $\Pi_{\mu}(p,p',q)$ can be decomposed as
\begin{eqnarray}
\notag
\Pi^{HAD}_{\mu}(p,p',q)=&&i^{2}\frac{Cg_{\Sigma_{c}^{*}ND[\Sigma_{b}^{*}NB]}}{(p^{2}-m^{2}_{\Sigma_{c}^{*}[\Sigma_{b}^{*}]})(p'^{2}-m^{2}_{N})(q^{2}-m^{2}_{D[B]})}\\
 \notag
&&\times \Big\{\Pi_{1}q_{\mu}+\Pi_{2}p\!\!\!/q_{\mu}+\Pi_{3}q\!\!\!/q_{\mu}+\Pi_{4}q\!\!\!/p\!\!\!/q_{\mu} \\
 \notag &&+\Pi_{5}p_{\mu}+\Pi_{6}p\!\!\!/p_{\mu}+\Pi_{7}q\!\!\!/p_{\mu}+\Pi_{8}q\!\!\!/p\!\!\!/p_{\mu}  \\
 \notag
&&+\Pi_{9}\gamma_{\mu}+\Pi_{10}p\!\!\!/\gamma_{\mu}+\Pi_{11}q\!\!\!/\gamma_{\mu}+\Pi_{12}q\!\!\!/p\!\!\!/\gamma_{\mu}
\Big\} \\
&&+\cdots.
\end{eqnarray}
Where
$C=\frac{m_{D[B]}^{2}f_{D[B]}\lambda_{N}\lambda_{\Sigma_{c}^{*}[\Sigma_{b}^{*}]}}{m_{c[b]}+m_{u}}$,
and some of the Dirac structure appearing in the above function are
written as
\begin{eqnarray}
\notag
&&\Pi_{1}=(m_{\Sigma_{c}^{*}[\Sigma_{b}^{*}]}+m_{N})m_{\Sigma_{c}^{*}[\Sigma_{b}^{*}]},\\
\notag
&&\Pi_{2}=(m_{\Sigma_{c}^{*}[\Sigma_{b}^{*}]}+m_{N}),\\
\notag &&\Pi_{i}=\cdots\cdots\cdots\cdots,\\
&&
\Pi_{12}=\frac{m_{\Sigma_{c}^{*}[\Sigma_{b}^{*}]}^{2}+2m_{\Sigma_{c}^{*}[\Sigma_{b}^{*}]}m_{N}+m_{N}^{2}-q^{2}}{6m_{\Sigma_{c}^{*}[\Sigma_{b}^{*}]}}.
\end{eqnarray}
In our previous analysis about this kind of problem, we found that
the Dirac structure $q\!\!\!/p\!\!\!/\gamma_{\mu}$ can not lead to
contaminations for $\Sigma_{c}^{*}[\Sigma_{b}^{*}]$\cite{WangZG5}.
Thus, we choose the Dirac structure $q\!\!\!/p\!\!\!/\gamma_{\mu}$
to carry out our analysis. In this above derivation, we also use the
following definitions,
\begin{eqnarray}
\notag
&&\Sigma u_{N}(p',s')\overline{u}_{N}(p',s')=p\!\!\!/'+m_{N},  \\
&&\Sigma
u_{\alpha\Sigma_{c}^{*}[\Sigma_{b}^{*}]}(p,s)\overline{u}_{\mu\Sigma_{c}^{*}[\Sigma_{b}^{*}]}(p,s)=-(p\!\!\!/+m_{\Sigma_{c}^{*}[\Sigma_{b}^{*}]})
(g_{\alpha\mu}-\frac{\gamma_{\alpha}\gamma_{\mu}}{3m_{\Sigma_{c}^{*}[\Sigma_{b}^{*}]}}-\frac{2p_{\alpha}p_{\mu}}{3m^{2}_{\Sigma_{c}^{*}[\Sigma_{b}^{*}]}}+
\frac{p_{\alpha}\gamma_{\mu}-\gamma_{\alpha}
p_{\mu}}{3m_{\Sigma_{c}^{*}[\Sigma_{b}^{*}]}})
\end{eqnarray}

\begin{large}
\textbf{2.2 The OPE side}
\end{large}

We carry out the operator product expansion of the correlation
function in deep Euclidean region, where $p^2\rightarrow-\infty$ and
$p'^2\rightarrow-\infty$. Considering all possible contractions of
the quark fields with Wick's theorem, the correlation function
Eq.$(1)$ is written as
\begin{eqnarray}
\notag\ \Pi_{\mu}^{OPE}(p,p',q)=&&i^{2}\int d^{4}x\int
d^{4}ye^{-ip.x}e^{ip'.y}\epsilon_{abc}\epsilon_{ijk}\\
&& \notag
\times\Big\{\gamma_{5}\gamma_{\nu}S_{d}^{cj}(y-x)\gamma_{\mu}CS_{u}^{biT}(y-x)C\gamma_{\nu}S_{u}^{ah}(y)\gamma_{5}S_{c[b]}^{hk}(-x)\\
&&-\gamma_{5}\gamma_{\nu}S_{d}^{cj}(y-x)\gamma_{\mu}CS_{u}^{aiT}(y-x)C\gamma_{\nu}S_{u}^{bh}(y)\gamma_{5}S_{c[b]}^{hk}(-x)\Big\}
\end{eqnarray}
Then, we replace the $c[b]$ and $u[d]$ quark propagators with the
corresponding full propagators\cite{Pasc,Wzg4,Rein},
\begin{eqnarray}
\notag
 S_{c[b]}^{mn}(x)=&&\frac{i}{(2\pi)^{4}}\int
 d^{4}ke^{-ik.x}\Big\{\frac{\delta_{mn}}{k\!\!\!/-m_{b[c]}}-\frac{g_{s}G_{mn}^{\alpha\beta}}{4}\frac{\sigma_{\alpha\beta}(k\!\!\!/+m_{b[c]})+(k\!\!\!/+m_{b[c]})\sigma_{\alpha\beta}}{(k^{2}-m_{Q}^{2})^{2}}\\
 &&+\frac{\pi^{2}}{3}\Big\langle\frac{\alpha_{s}GG}{\pi}\Big\rangle\delta_{mn}m_{b[c]}\frac{k^{2}+m_{b[c]}k\!\!\!/}{(k^{2}-m_{b[c]}^{2})^{4}}+\cdots\Big\},
\end{eqnarray}
\begin{eqnarray}
 S_{u[d]}^{mn}(x)=&&i\frac{x\!\!\!/}{2\pi^{2}x^{4}}\delta_{mn}-\frac{m_{u[d]}}{4\pi^{2}x^{2}}\delta_{mn}-\frac{\langle
 \overline{q}q\rangle}{12}\Big(1-i\frac{m_{u[d]}}{4}x\!\!\!/\Big)-\frac{x^{2}}{192}m_{0}^{2}\langle
 \overline{q}q\rangle\Big( 1-i\frac{m_{u[d]}}{6}x\!\!\!/\Big)\\
 &&
 \notag-\frac{ig_{s}\lambda_{A}^{ij}G^{A}_{\theta\eta}}{32\pi^{2}x^{2}}\Big[x\!\!\!/\sigma^{\theta\eta}+\sigma^{\theta\eta}x\!\!\!/\Big]+\cdots.
\end{eqnarray}
In order to perform the four-$x$ and four-$y$ integrals, we also use
the following Fourier transformations in $D=4+\epsilon$ dimensions
with $\epsilon\rightarrow0$,
\begin{eqnarray}
\frac{1}{[(y-x)^{2}]^{n}}=\int\frac{d^{D}t}{(2\pi)^{D}}e^{-it.(y-x)}i(-1)^{n+1}2^{D-2n}\pi^{D/2}\frac{\Gamma(D/2-n)}{\Gamma(n)}\Big(-\frac{1}{t^{2}}\Big)^{D/2-n},
\end{eqnarray}
\begin{eqnarray}
\frac{1}{[y^{2}]^{n}}=\int\frac{d^{D}t'}{(2\pi)^{D}}e^{-it'.y}i(-1)^{n+1}2^{D-2n}\pi^{D/2}\frac{\Gamma(D/2-n)}{\Gamma(n)}\Big(-\frac{1}{t'^{2}}\Big)^{D/2-n},
\end{eqnarray}
which is followed by the replacements $x_{\mu}\rightarrow
i\frac{\partial}{\partial p_{\mu}}$ and $y_{\mu}\rightarrow
-i\frac{\partial}{\partial p'_{\mu}}$. After these derivations,
these integrals turn into Dirac delta functions which are used to
take the four-integrals over $k$ and $t'$. Finally, the Feynman
parametrization and
\begin{eqnarray}
\notag\
\int\frac{d^{D}t}{(2\pi)^{D}}\frac{1}{[t-M^{2}]^{\alpha}}&&=\frac{i(-1)^{\alpha}}{(4\pi)^{D/2}}\frac{\Gamma(\alpha-D/2)}{\Gamma(\alpha)}\frac{1}{(M^{2})^{\alpha-D/2}}
\end{eqnarray}
is used to perform the four-$t$ integral. After a lengthy
derivation, we obtain the same Dirac structures as the
phenomenological side(see Eq.$(5)$). For each Dirac structure, the
correlation function can be divided into two parts,
\begin{eqnarray}
\Pi^{OPE}_{i}=\Pi^{pert}_{i}+\Pi^{non-pert}_{i}
\end{eqnarray}
Where $i$ stands for different Dirac structure in Eq.$(5)$. Using
dispersion relations, the perturbative term for a given Dirac
structure can be written as the following form,
\begin{eqnarray}
\Pi^{pert}_{i}(q^{2})=\int_{s_{1}}^{s_{0}} ds\int_{u_{1}}^{u_{0}}
du\frac{\rho_{i}^{pert}(s,u,q^{2})}{(s-p^{2})(u-p'^{2})}
\end{eqnarray}
where $\rho_{i}^{pert}(s,u,q^{2})$, appearing in the above equation,
is the spectral density which is obtained from the imaginary part of
the correlation. After these derivations, we set $s=p^2$, $u=p'^2$
and $q=p-p'$ in the spectral densities. For the Dirac structure of
$q\!\!\!/p\!\!\!/\gamma_{\mu}$, its spectral density is written as,
\begin{eqnarray}
\notag\
\rho_{q\!\!\!/p\!\!\!/\gamma_{\mu}}^{pert}(s,u,q^2)&&=\frac{3}{32\pi^{4}}\int^{1}_{0}dx\int^{1-x}_{0}\Big\{\frac{x+y}{2(x+y-1)}\Big[x(-q^2y+m_{c[b]^2})-sx(x+y-1)-uy(x+y-1)\Big] \\
\ &&
+\frac{1}{6}\Big[sx(x+y)+uy(x+y)+q^2xy-6m_{d}m_{u}\Big]\Big\}dy\times\Theta[H_{2}(s,u,q^{2})],
\end{eqnarray}
where $\Theta$ stands for the unit-step function, and
$H_{2}(s,u,q^2)$ is defined as $H_{2}[s,u,q^2]=x (m_{c[b]}^2-q^2
y)+s x (x+y-1)+u y (x+y-1)$. Considering the limit of the
unit-function to the integrals, the integral limmits for parameter
$y$ can be explicitly expressed as
\begin{eqnarray}
0\leq y\leq
min\Big\{1-x,\frac{[(s+u-q^2)-u]-\sqrt{\Delta}}{-2u}\Big\}
\end{eqnarray}
where
$\Delta=\Big[(s+u-q^2)^2-4su\Big]x^2-2u\Big[(s+u-q^2)+2m_{c[b]}^{2}\Big]x+4su+u^{2}$.

From our previous analysis, the non-perturbative contributions comes
mainly from the $<\overline{q}q>$. Besides of this condensate term,
we also take into account the contribution from
$<\frac{\alpha_{s}}{\pi}GG>$ in this work. For these condensate
terms, we make the change of variables $p^2\rightarrow-P^2$,
$p'^2\rightarrow-P'^2$ and $q^2\rightarrow-Q^2$ and perform a double
Borel transform to them, which involves the transformation:
$P^2\rightarrow M_{1}^{2}$ and $P'^2\rightarrow M_{2}^{2}$, where
$M_1$ and $M_2$ are the Borel parameters. Then, the non-perturbative
terms can be written as,
\begin{eqnarray}
\notag\
\mathscr{B}\mathscr{B}\Big[\Pi_{q\!\!\!/p\!\!\!/\gamma_{\mu}}^{<\overline{q}q>}(Q^2)\Big]&&=\frac{<\overline{q}q>}{64\pi^{2}}\int^{1}_{0}dx\int^{1-x}_{0}\Big\{\frac{4\Big[(12m_{d}+3m_{u})(x+y-1)+14m_{d}+5m_{u}\Big]}{x+y} \\
\notag\
&&+(4m_{u}+16m_{d})\Big[1+\frac{x(Q^2y+m_{c[b]}^{2})}{x(x+y-1)M_{1}^{2}}-\frac{y}{x+y}\frac{Q^2}{M_{1}^{2}}\Big]\Big\}dy
\\
\notag &&\times
\frac{e^{\frac{Q^2y+m_{c[b]}^{2}}{(x+y-1)M_{1}^{2}}}}{M_{1}^{2}+M_{2}^{2}}\delta\Big\{k-\frac{M_{2}^{2}}{M_{1}^{2}+M_{2}^{2}}\Big\},
\\
\end{eqnarray}
\begin{eqnarray}
\notag\
\mathscr{B}\mathscr{B}\Big[\Pi_{q\!\!\!/p\!\!\!/\gamma_{\mu}}^{<\overline{G}G>}(Q^2)\Big]&&=\frac{<\frac{\alpha_{s}}{\pi}GG>}{32\pi^{2}\times48}\int^{1}_{0}dx\int^{1-x}_{0}
\Big\{\frac{x^{2}}{6(x+y-1)^{3}}\Big[8+\frac{2x(Q^2y+m_{c[b]}^{2})}{x(x+y-1)M_{1}^{2}}-\frac{y}{x+y}\frac{Q^2}{M_{1}^{2}}\Big]
\\
\notag && \times\frac{m_{c[b]}^{2}}{M_{1}^{2}}\Big\}dy
\frac{e^{\frac{Q^2y+m_{c[b]}^{2}}{(x+y-1)M_{1}^{2}}}}{M_{1}^{2}+M_{2}^{2}}\delta\Big\{k-\frac{M_{2}^{2}}{M_{1}^{2}+M_{2}^{2}}\Big\},
\\
\end{eqnarray}
where $\mathscr{B}\mathscr{B}[$ $]$ stands for the double Borel
transform, $\delta$ is the Delta function and $k=\frac{x}{x+y}$.

\begin{large}
\textbf{2.3 The strong coupling constant}
\end{large}

\begin{large}
\textbf{3 The results and discussions}
\end{large}

We perform a double Borel transform to the phenomenological side as
well as the OPE side, after which we equate these two sides,
invoking the quark-hadron duality. After these preformation, the
form factor can be written as,
\begin{eqnarray}
\notag &&
g_{\Sigma_{c}^{*}ND(\Sigma_{b}^{*}NB)}(Q^{2}) \\
&&=\frac{\frac{1}{M_{1}^{2}M^{2}_{2}}\int_{(m_{c[b]}+m_{u}+m_{d})^2}^{s_{0}}
\int_{(2m_{u}+m_{d})^2}^{u_{0}}
\rho_{i}^{pert}(s,u,Q^{2})e^{-\frac{s}{M_{1}^{2}}}e^{-\frac{u}{M_{2}^{2}}}dsdu
+\mathscr{B}\mathscr{B}\Big[\Pi^{non-pert}(Q^2)\Big]}{\frac{C\Pi_{12}}{(Q^2+m_{D[B]}^2)M_1^2M_2^2}e^{-\frac{m_{\Sigma_{c}^*[\Sigma_{b}^*]}^2}{M_1^2}}e^{-\frac{m_N^2}{M_2^2}}}
\end{eqnarray}
Where $\Pi_{12}$ in the above equation represents the
$q\!\!\!/p\!\!\!/\gamma_{\mu}$ term in the phenomenological side in
Eq.$(5)$, $s_{0}$ and $u_{0}$ are the continuum threshold parameters
which are used to eliminate the $h.r.$ terms. Commonly, the
continuum parameters, $s_{0}=(m_{i}+\Delta_{i})^2$ and
$u_{0}=(m_{o}+\Delta_{o})^2$ are employed to include the pole and
suppress the $h.r.$ contributions, where $m_{i}$ and $m_{o}$ are the
masses of the incoming and out-coming baryons respectively. In
general, $\Delta_{i}$ and $\Delta_{o}$ are chosen to be about
$0.5GeV^2$ for mesons, whose value is some what smaller than that of
the baryons.
\begin{figure}[h]
\begin{minipage}[t]{0.45\linewidth}
\centering
\includegraphics[height=5cm,width=7cm]{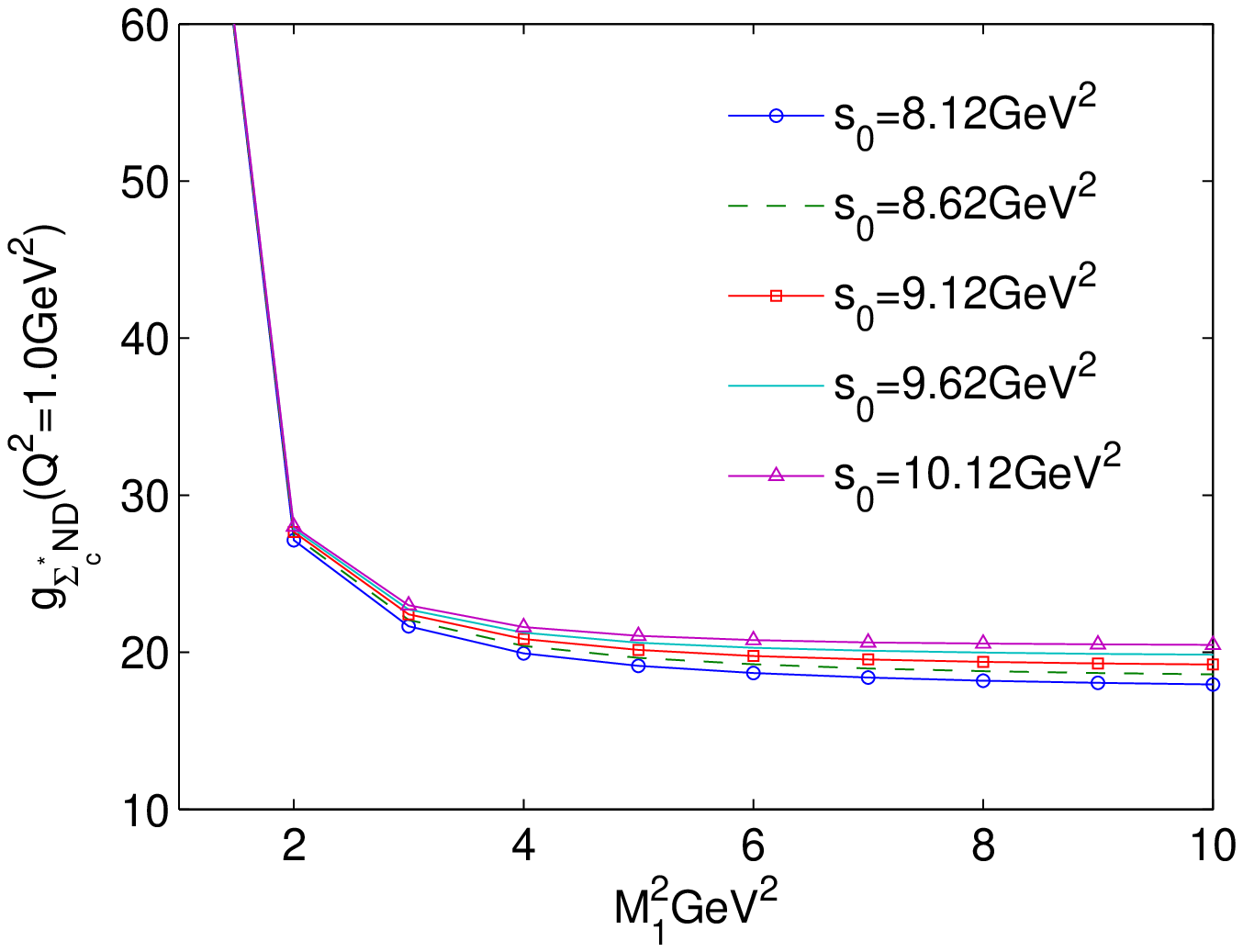}
\caption{The strong form factor $g_{\Sigma_{c}^{*}ND}$ on Borel
parameter $M_1^2$, in the different values of $s_{0}$.\label{your
label}}
\end{minipage}
\hfill
\begin{minipage}[t]{0.45\linewidth}
\centering
\includegraphics[height=5cm,width=7cm]{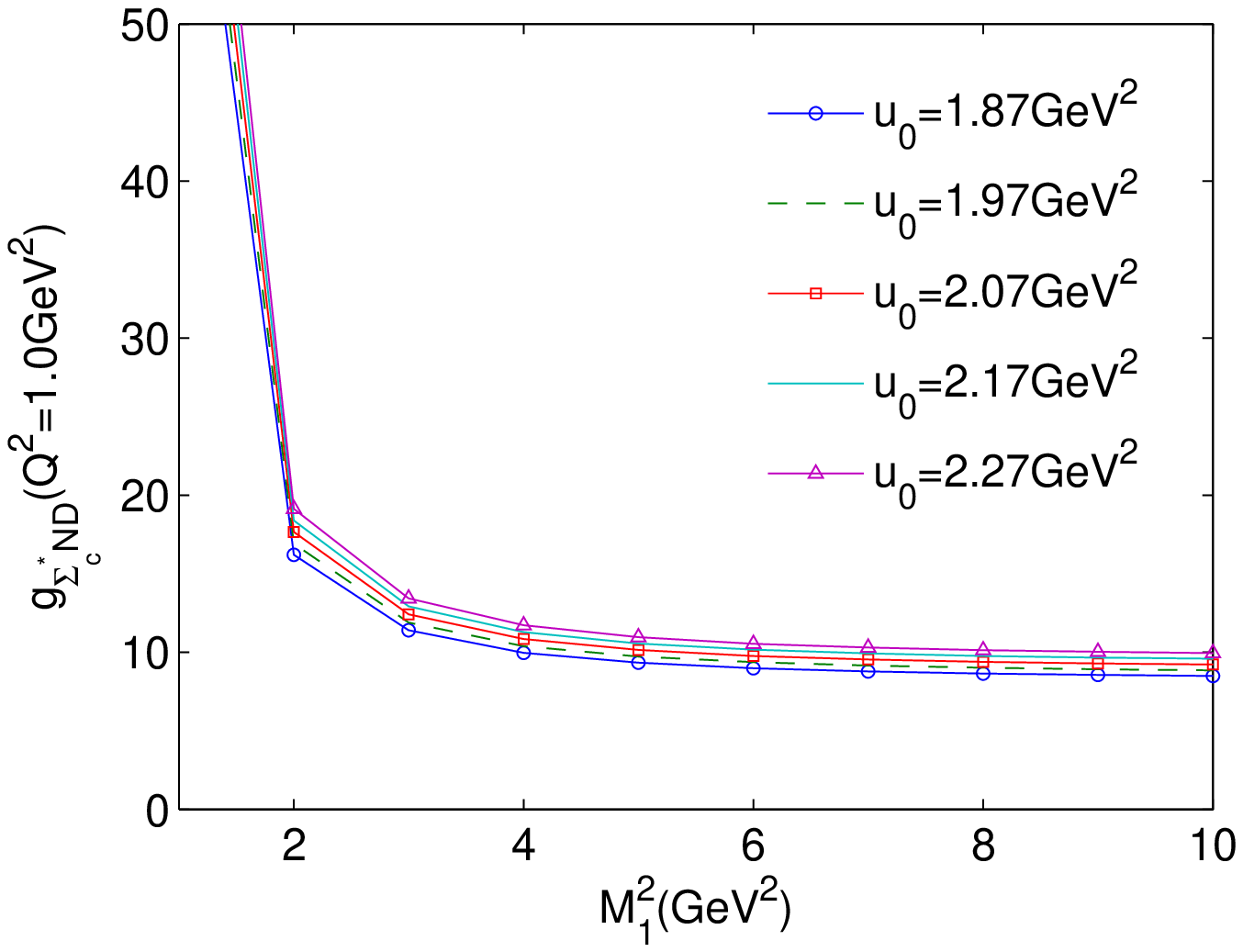}
\caption{The strong form factor $g_{\Sigma_{c}^{*}ND}$ on Borel
parameter $M_1^2$, in the different values of $u_0$.\label{your
label}}
\end{minipage}
\end{figure}

It can also be seen from Eq.$(19)$ that the form factor
$g_{\Sigma_{c}^{*}ND[\Sigma_{b}^{*}NB]}$ is the function of the
Borel parameters $M_{1}^{2}$ and $M_{2}^{2}$. We determine the
working regions for the Borel parameters according to two
considerations which are pole dominance and convergence of the OPE.
That is to say, the pole contribution should be as large as possible
comparing with the contributions of the higher and continuum states.
Meanwhile, we should also find a plateau, which will ensure OPE
convergence and the stability of our results. The plateau is often
called "Borel window".

The form factor, $g_{\Sigma_{c}^{*}ND}$, on Borel parameter $M_1^2$
in the different values of $s_{0}$ and $u_0$ are shown in Figs.1 and
2, where $9.12GeV^2\approx(m_{\Sigma_{c}^{*}}+0.5GeV)^2$ and
$2.07GeV^2\approx(m_{D}+0.5GeV)^2$ in these figures. It can be seen
from Fig.1 that the value of $g_{\Sigma_{c}^{*}ND}$ show more
stability with $s_{0}=10.12GeV^2$. However, we can see from Fig.2
that the line of $g_{\Sigma_{c}^{*}ND}$ show little change with
$u_{0}$ changes from $1.87GeV^2$ to $2.27GeV^2$. Finally, the
continuum threshold parameters are chosen to be $u_{0}=2.07GeV^2$
and $s_{0}=10.12GeV^2[39.00GeV^2]$ for vertex
$\Sigma_{c}^{*}ND[\Sigma_{b}^{*}NB]$. In Figs.$3-6$, we show also
the relative continuum and pole contribution on Borel parameter. We
can see from these figures that the more little values of the Borel
parameters lead to larger pole contributions in the results.
However, if too little values of the Borel parameters are taken(see
Figs.1 and 2), the results decrease monotonously and quickly with
the Borel parameters, which means that the convergence of the OPE
can not be satisfied. Finally, the Borel windows are chosen to be
$3.0GeV^{2}<M_{1}^{2}<5.0GeV^{2}$ and
$1.0GeV^{2}<M_{2}^{2}<3.0GeV^{2}$ for the vertex $\Sigma_{c}^{*}ND$,
 and $9.0GeV^{2}<M_{1}^{2}<11.0GeV^{2}$ and
$1.0GeV^{2}<M_{2}^{2}<3.0GeV^{2}$ for the vertex $\Sigma_{b}^{*}NB$.
Under these circumstances the criteria of pole dominance and OPE
convergence are all satisfied. It can also be seen from Figs.7-9
that our results have weak dependence on the Borel parameters, which
indicates the stability of the results.
\begin{figure}[h]
\begin{minipage}[t]{0.45\linewidth}
\centering
\includegraphics[height=5cm,width=7cm]{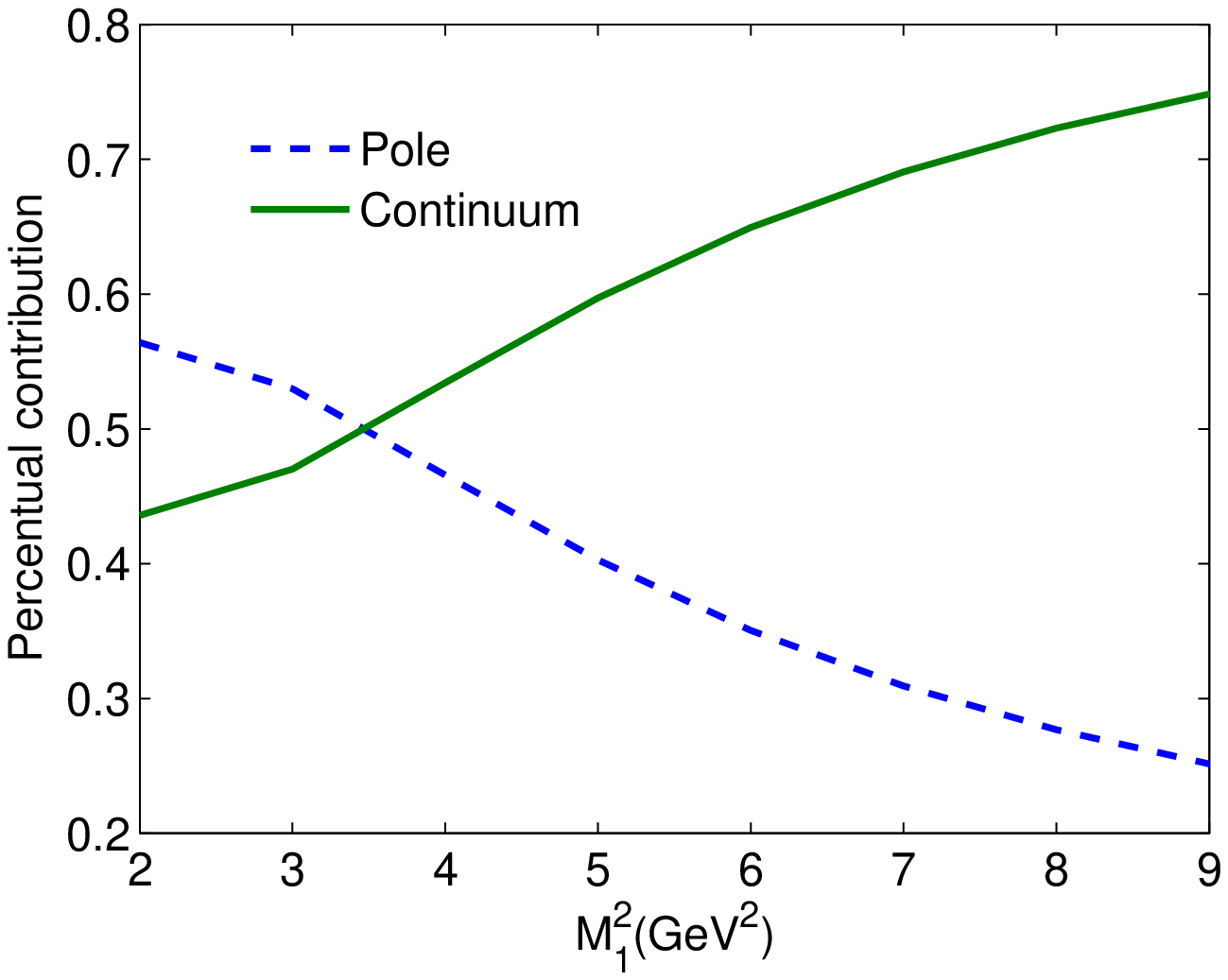}
\caption{Pole and continuum contributions for vertex $\Sigma_{c}^{*}
ND$, on Borel parameter $M_1^2$.\label{your label}}
\end{minipage}
\hfill
\begin{minipage}[t]{0.45\linewidth}
\centering
\includegraphics[height=5cm,width=7cm]{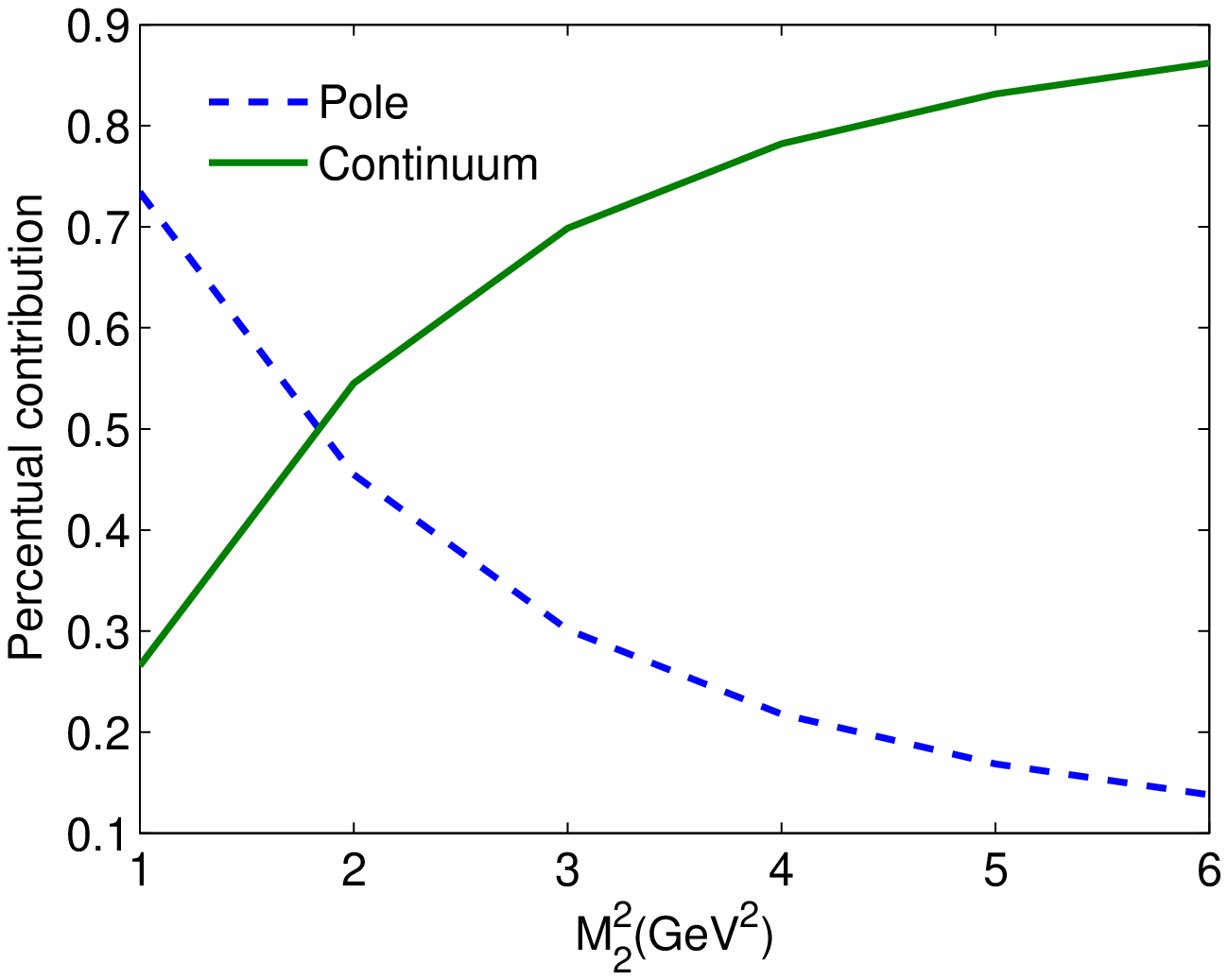}
\caption{Pole and continuum contributions for vertex $\Sigma_{c}^{*}
ND$, on Borel parameter $M_2^2$.\label{your label}}
\end{minipage}
\end{figure}
\begin{figure}[h]
\begin{minipage}[t]{0.45\linewidth}
\centering
\includegraphics[height=5cm,width=7cm]{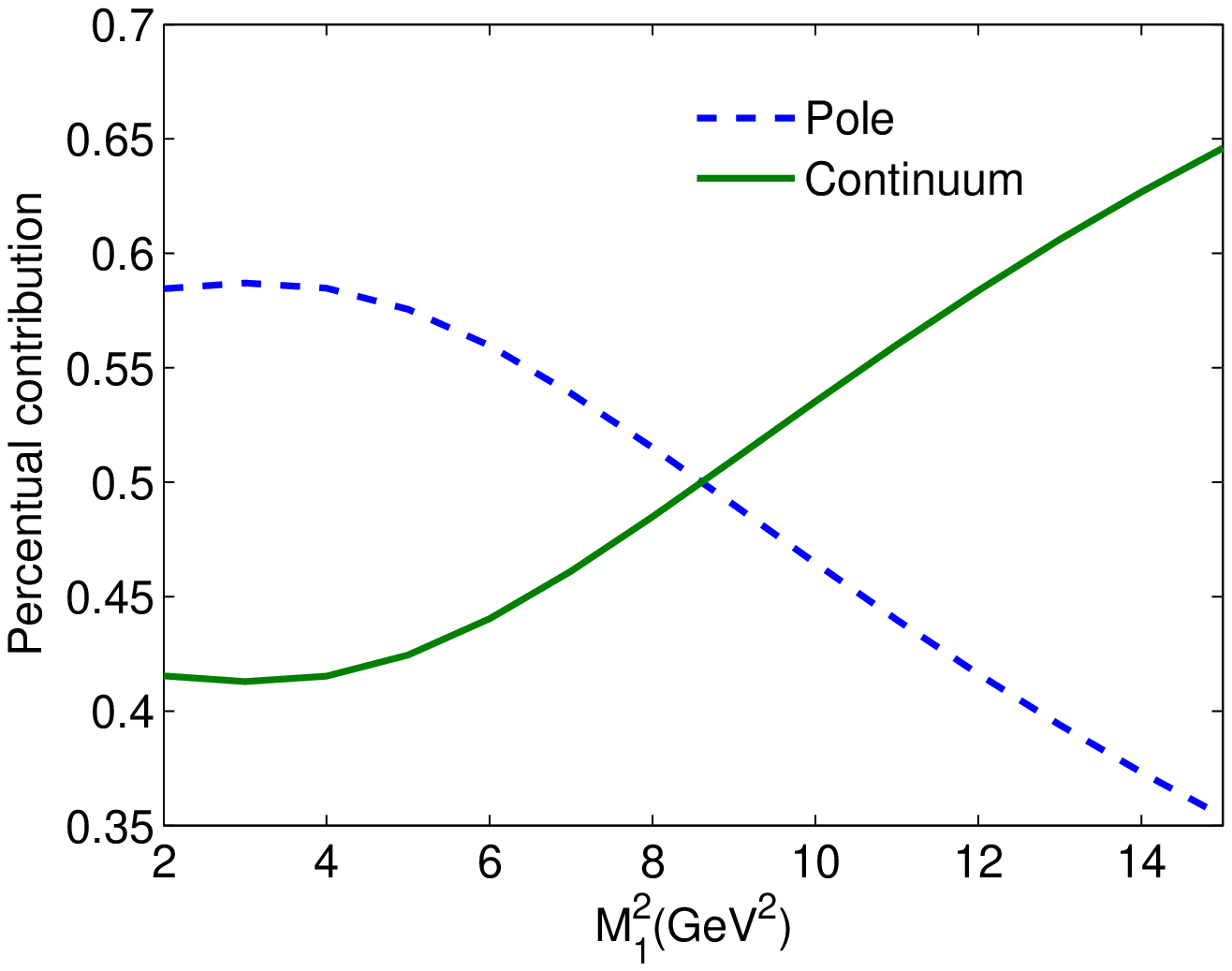}
\caption{Pole and continuum contributions for vertex $\Sigma_{b}^{*}
NB$, on Borel parameter $M_1^2$.\label{your label}}
\end{minipage}
\hfill
\begin{minipage}[t]{0.45\linewidth}
\centering
\includegraphics[height=5cm,width=7cm]{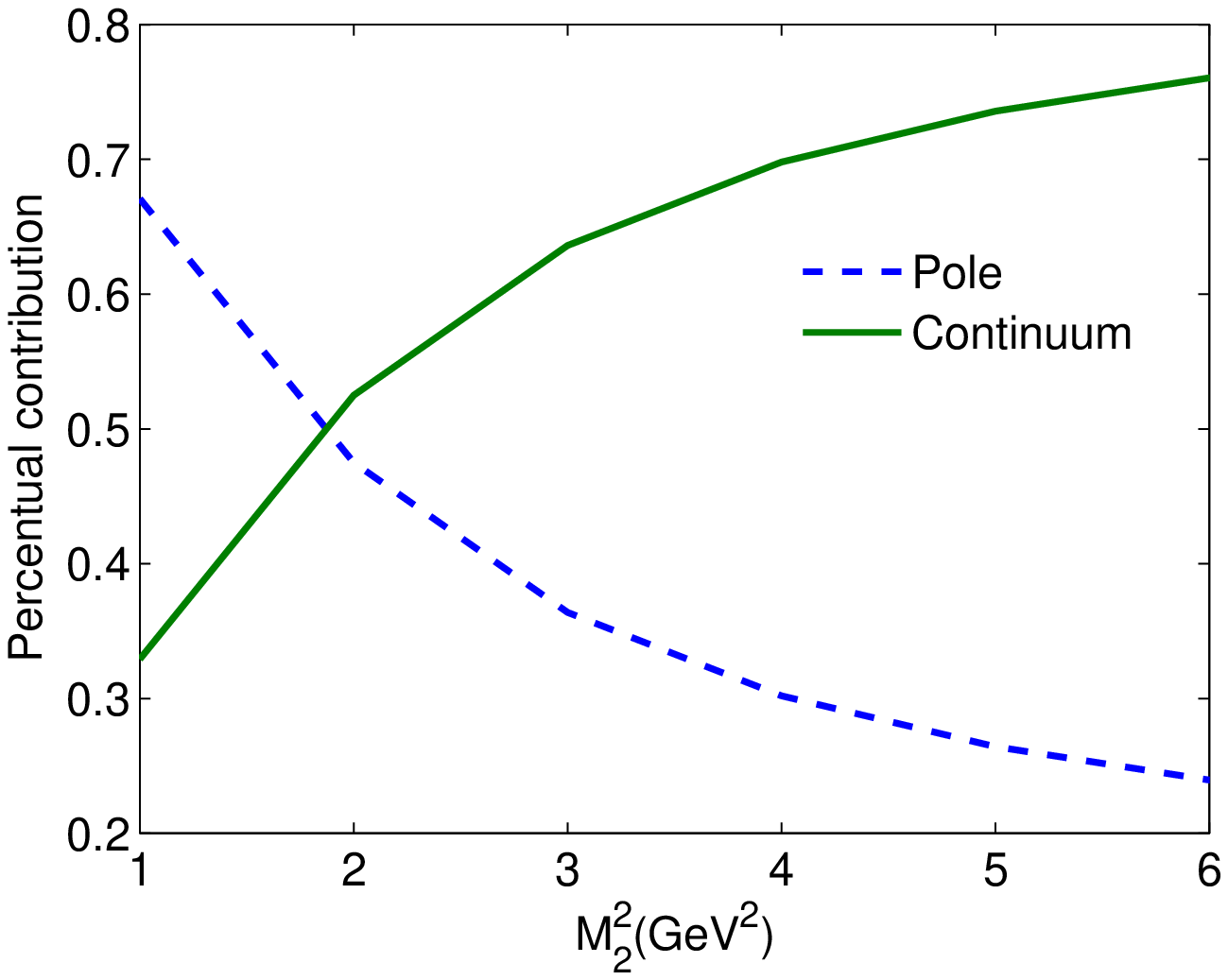}
\caption{Pole and continuum contributions for vertex $\Sigma_{b}^{*}
NB$, on Borel parameter $M_2^2$\label{your label}}
\end{minipage}
\end{figure}
\begin{figure}[h]
\begin{minipage}[t]{0.45\linewidth}
\centering
\includegraphics[height=5cm,width=7cm]{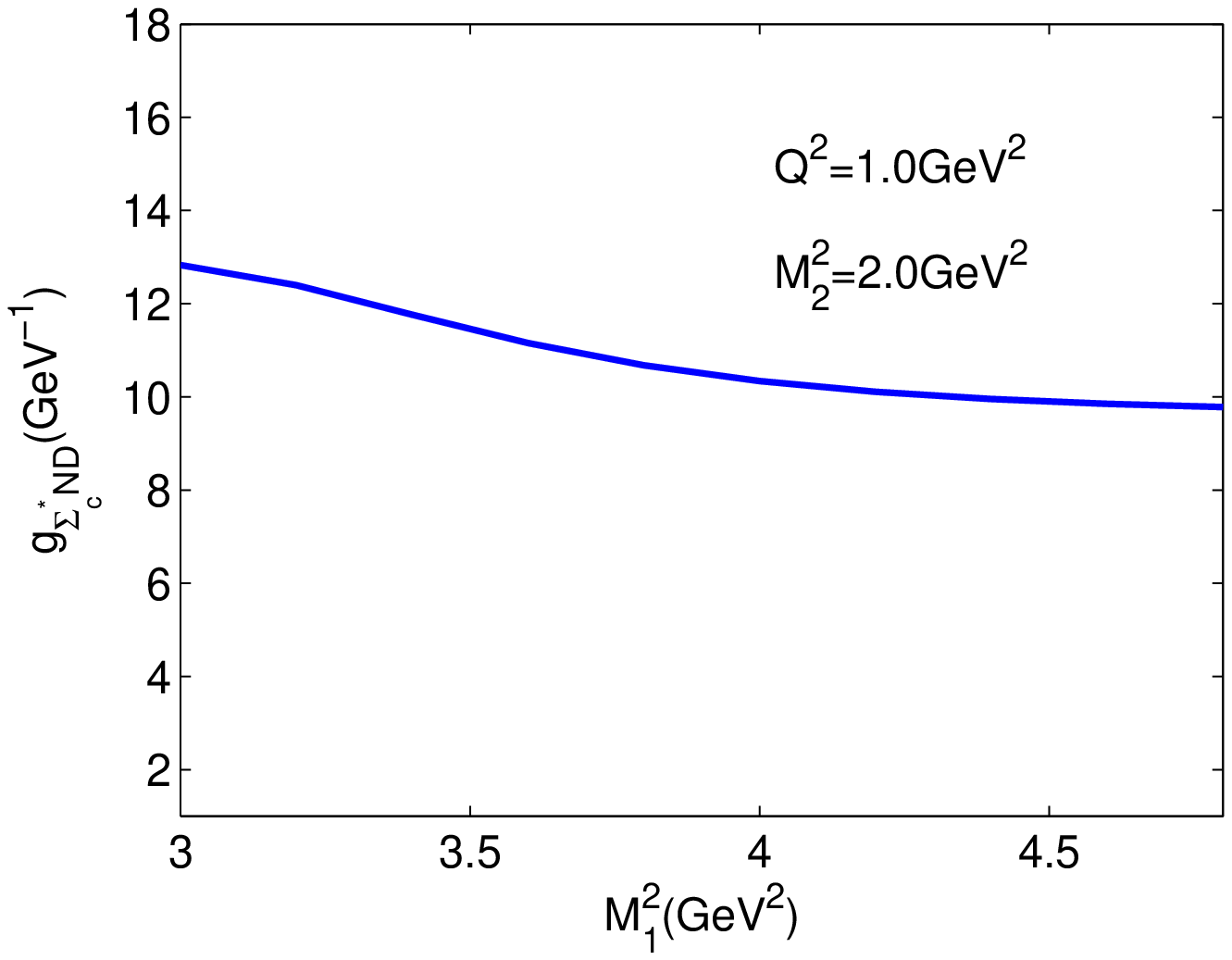}
\caption{$g_{\Sigma_{c}^{*}ND}$ ($Q^2=1.0GeV^2$) as a function of
the Borel mass $M_{1}^{2}$.\label{your label}}
\end{minipage}
\hfill
\begin{minipage}[t]{0.45\linewidth}
\centering
\includegraphics[height=5cm,width=7cm]{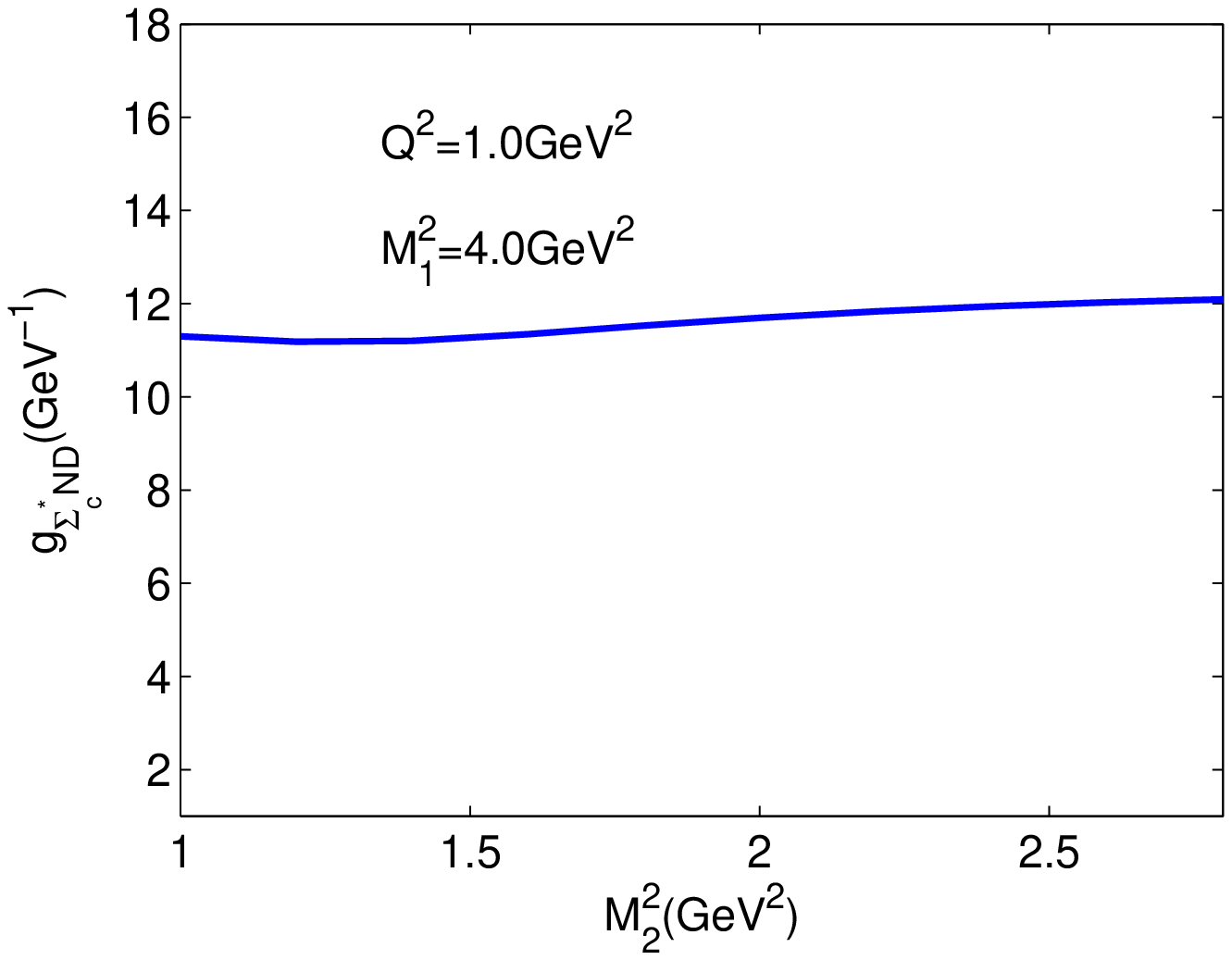}
\caption{$g_{\Sigma_{c}^{*}ND}$ ($Q^2=1.0GeV^2$) as a function of
the Borel mass $M_{2}^{2}$\label{your label}}
\end{minipage}
\end{figure}
\begin{figure}[h]
\begin{minipage}[t]{0.45\linewidth}
\centering
\includegraphics[height=5cm,width=7cm]{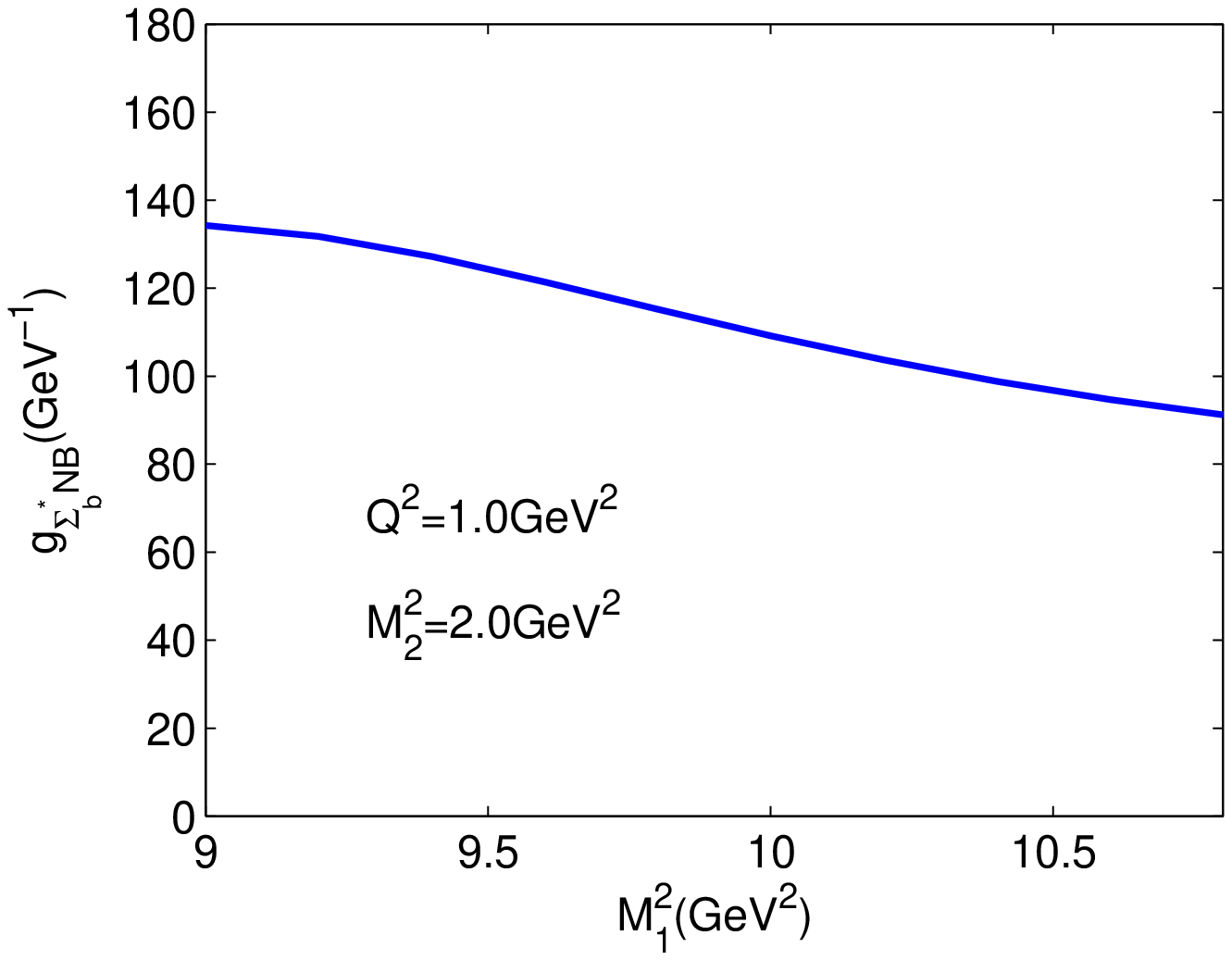}
\caption{$g_{\Sigma_{b}^{*}NB}$ ($Q^2=1.0GeV^2$) as a function of
the Borel mass $M_{1}^{2}$\label{your label}}
\end{minipage}
\hfill
\begin{minipage}[t]{0.45\linewidth}
\centering
\includegraphics[height=5cm,width=7cm]{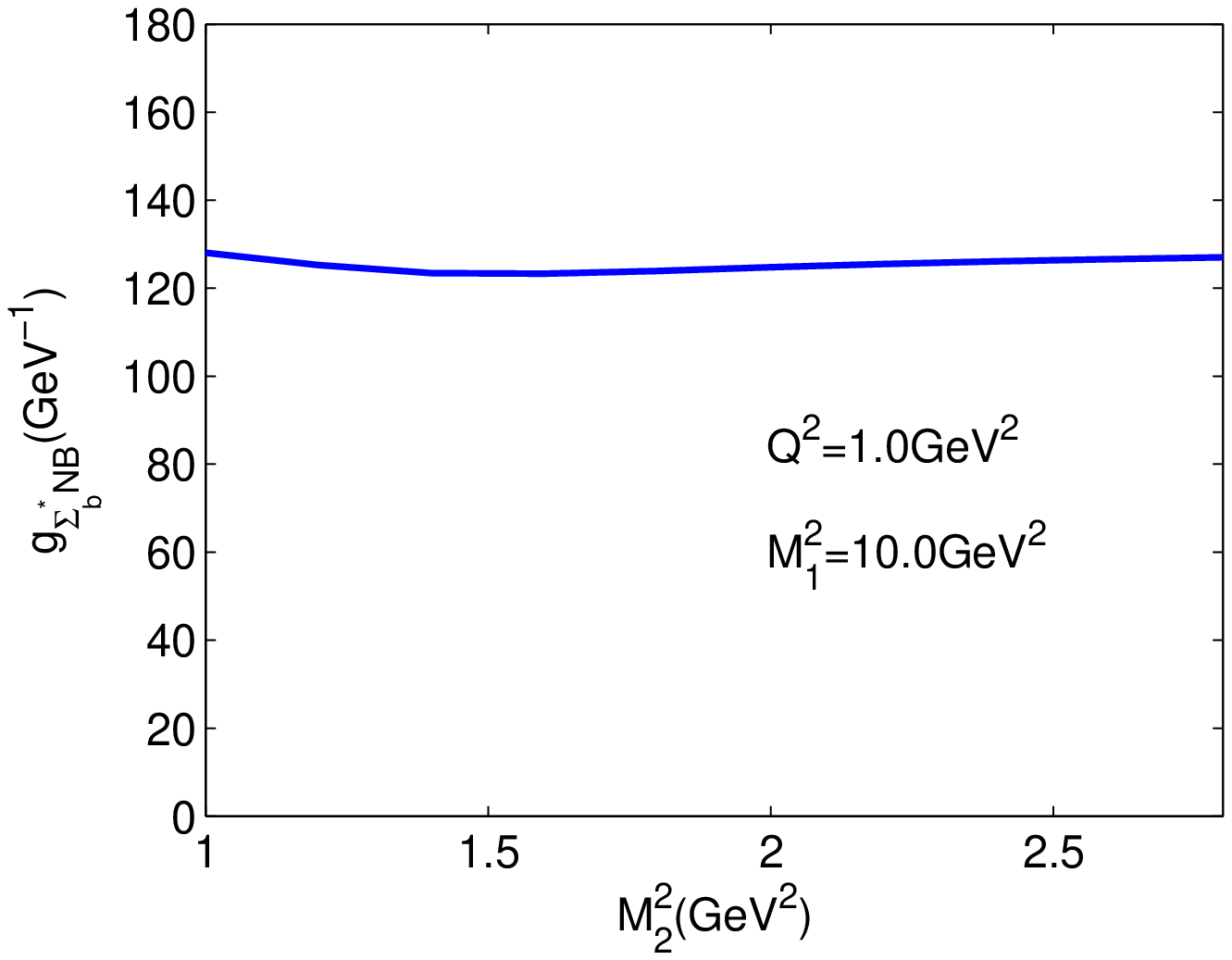}
\caption{$g_{\Sigma_{b}^{*}NB}$ ($Q^2=1.0GeV^2$) as a function of
the Borel mass $M_{2}^{2}$\label{your label}}
\end{minipage}
\end{figure}
\begin{figure}[h]
\begin{minipage}[t]{0.45\linewidth}
\centering
\includegraphics[height=5cm,width=7cm]{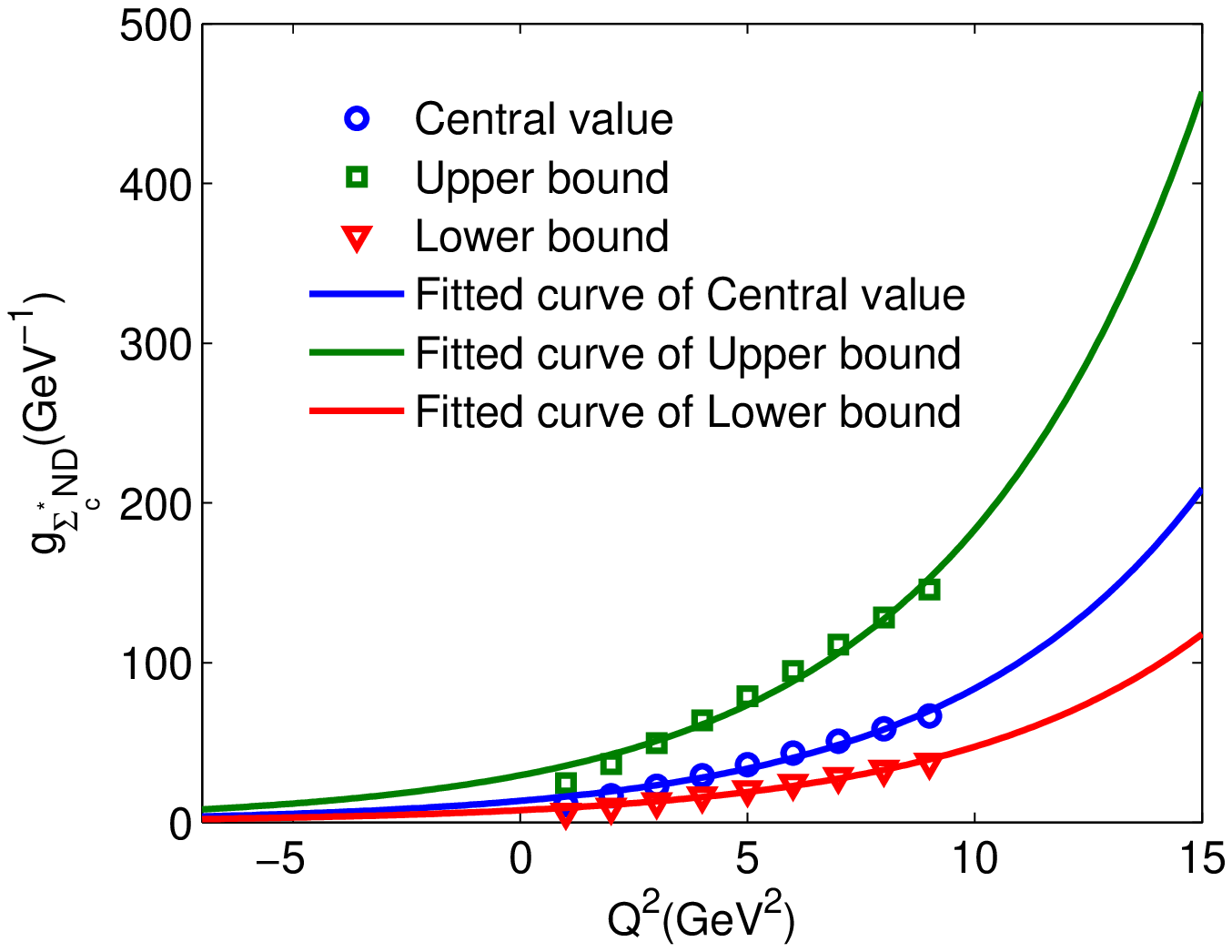}
\caption{The form factor for the vertex $\Sigma_{c}^{*}ND$, and its
fitted results as a function of $Q^2$.\label{your label}}
\end{minipage}
\hfill
\begin{minipage}[t]{0.45\linewidth}
\centering
\includegraphics[height=5cm,width=7cm]{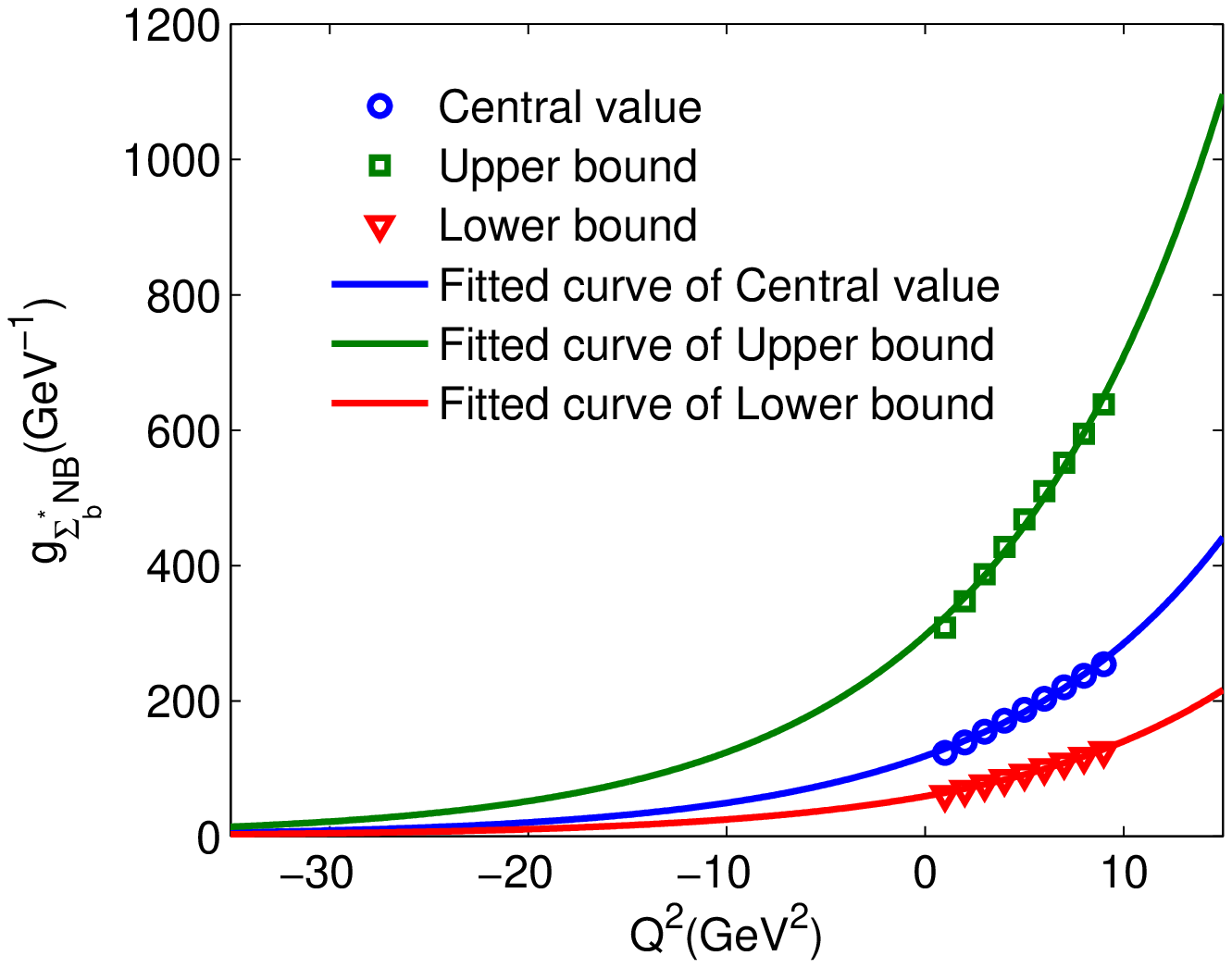}
\caption{The form factor for the vertex $\Sigma_{b}^{*}NB$, and its
fitted results as a function of $Q^2$.\label{your label}}
\end{minipage}
\end{figure}
\begin{table*}[t]
\begin{ruledtabular}\caption{Input parameters used in this analysis.}
\begin{tabular}{c c }
  Parameters & \ Values   \\
\hline
$m_{\Sigma_{b}^{*}}$    &  \   $5835.1\pm1.9$  $MeV$ \cite{Oliv}  \\
$m_{\Sigma_{c}^{*}}$     &  \    $2517.5\pm2.3$  $MeV$ \cite{Oliv}       \\
$m_{N}$       &  \   $ 939.565379 \pm 0.000021$  $MeV$  \cite{Oliv}        \\
$m_{D}$       &  \      $1864.86 \pm 0.13$  $MeV$  \cite{Oliv}    \\
$m_{B}$      &  \   $5279.58 \pm 0.17$    $MeV$  \cite{Oliv}    \\
$m_{b}$       &  \     $4.18\pm0.03$ $GeV$ \cite{Oliv}      \\
$m_{c}$          &  \    $1.275 \pm 0.025$   $GeV$ \cite{Oliv}   \\
$m_{d}$        &  \    $4.8^{+0.5}_{-0.3}$ $MeV$  \cite{Oliv}       \\
$m_{u}$        &  \    $2.3^{+0.7}_{-0.5}$ $MeV$   \cite{Oliv}      \\
$\langle \overline{u}u\rangle$         &  \  $-(0.24\pm0.01)GeV^3$\cite{Ioff2}           \\
$\langle \overline{d}d\rangle$        &  \  $-(0.24\pm0.01)GeV^3$  \cite{Ioff2}           \\
$\langle \frac{\alpha_{s}G^{2}}{\pi}\rangle$        &  \    $(0.012\pm0.04)$  $GeV^4$  \cite{Bely}       \\
$\lambda_{\Sigma_{b}^{*}}$        &  \     $0.0125^{+0.0008}_{-0.0004}GeV^3$ \cite{WangZG}        \\
$\lambda_{\Sigma_{c}^{*}}$        &  \    $0.0246^{+0.0022}_{-0.0016}GeV^3 $ \cite{WangZG}         \\
$\lambda_{N}^{2}$        &  \   $0.0011\pm0.0005$ $GeV^6$   \cite{Azi3}        \\
$f_{B}$        &  \    ($248\pm23_{exp}\pm25_{Vub}$)    $MeV$  \cite{Khod4}    \\
$f_{D}$        &  \    ($205.8\pm8.5\pm2.5$)    $MeV$  \cite{Eise}        \\
\end{tabular}
\end{ruledtabular}
\end{table*}

In order to obtain the coupling constants, it is necessary to
extrapolate these calculated results of the form factors into the
deep time-like regions by fitting these results into analytical
functions. The extrapolation to deep time-like regions is highly
mode-dependent, thus there is no specific expressions for the
dependence of the strong form factors on $Q^2$. From our analysis,
we observe that this dependence can be well described by the
following fit function
\begin{equation}
g_{\Sigma_{c}^{*}ND(\Sigma_{b}^{*}NB)}(Q^{2})=A_{c(b)}exp[{B_{c(b)}Q^2}]
\end{equation}
The fitted results for $g_{\Sigma_{c}^{*}ND}$ and
$g_{\Sigma_{b}^{*}NB}$ are $A_{c}=13.54\pm3.00$$GeV^{-1}$,
$B_{c}=0.1824\pm0.03$$GeV^{-2}$, and
$A_{b}=118.60\pm5.00$$GeV^{-1}$, $B_{b}=0.08675\pm0.006$$GeV^{-2}$.
In Figs.11 and 12, we show the dependence of the strong coupling
form factors on $Q^2$ for both the QCD sum rules and fitting
results, in which it is marked as Central value and fitted curve of
Central value. The values of the strong coupling constants can be
obtained from the fit function at $Q^{2}=-m_{B[D]}^2$, which are
$g_{\Sigma_{c}^{*}ND}=7.19\pm1.76$ and
$g_{\Sigma_{b}^{*}NB}=10.54\pm1.82$. The errors appearing in these
results are coming from the uncertainties of the fitting parameters
of $\delta A_{c}$, $\delta B_{c}$, $\delta A_{b}$ and $\delta
B_{b}$.

The uncertainties of the results coming from the input parameters
can be estimated with the formula
$\delta=\sqrt{\Sigma_{i}(\frac{\partial f}{\partial
x_{i}})^{2}(x_{i}-\overline{x}_{i})^{2}}$, where the $f$ denotes
strong form factors $g_{\Sigma_{c}^{*}ND}$ and
$g_{\Sigma_{b}^{*}NB}$, the $x_{i}$ denotes the revelent parameters
$m_{\Sigma_{b}^{*}}$,$m_{\Sigma_{c}^{*}}$,$m_{b}$,$m_{c}$,$\lambda_{\Sigma_{b}^{*}}$,
$\lambda_{\Sigma_{c}^{*}}$,$\langle \overline{q}q\rangle$,$\cdots$.
For simplicity, the value of the upper and lower limits of the
strong form factors $g_{\Sigma_{c}^{*}ND}$, $g_{\Sigma_{b}^{*}NB}$
are approximated by taking
$f^{upper(lower)}=f(\overline{x}_{i}\pm\Delta x_{i})$, which are
marked as Upper bound and Lower bound in Figures 11 and 12. After
these approximations, the results are also fitted into the same kind
of analytical function with Eq.(20) and are also extrapolated into
the physical regions in order to get the uncertainties of the
coupling constants. Finally, we get the strong coupling constants
for these two vertexes,
\begin{eqnarray}
g_{\Sigma_{c}^{*}ND}(Q^2=-m_{D}^{2})=7.19^{+8.49}_{-3.11}\pm1.76 \\
g_{\Sigma_{b}^{*}NB}(Q^2=-m_{B}^{2})=10.54^{+15.59}_{-5.23}\pm1.82
\end{eqnarray}
where the first part of the uncertainties in the results comes from
the input parameters,
$m_{\Sigma_{b}^{*}}$,$m_{\Sigma_{c}^{*}}$,$m_{b}$,$m_{c}$,$\lambda_{\Sigma_{b}^{*}}$,
$\lambda_{\Sigma_{c}^{*}}$,$\langle \overline{q}q\rangle$,$\cdots$
and the second part originates from the fitting parameters.

\begin{large}
\textbf{4 Conclusion}
\end{large}

In this article, we have calculated the form factors of the strong
vertexes $\Sigma_c^{*} ND$ and $\Sigma_b^{*}NB$ in the space-like
regions by three-point QCD sum rules. Then we fit the form factors
into analytical functions, extrapolate them into the time-like
regions, and obtain the strong coupling constants $g_{\Sigma_c^{*}
ND}$ and $g_{\Sigma_b^{*}NB}$. These results will be helpful in the
bottom and charmed meson cloud description of the nucleon, which may
be used to explain exotic events observed by different experiments.
Besides, the analysis of the results in heavy ion collision
experiments may also needs the results in this paper.


\begin{large}
\textbf{Acknowledgment}
\end{large}

This work has been supported by the Fundamental Research Funds for
the Central Universities, Grant Number $2016MS133$.


\begin{thebibliography}{99}

\bibitem{Aube06} B. Aubert $et$ $al$, Phys. Rev. Lett. 97, 232001 (2006)


\bibitem{Naka10} K. Nakamura $et$ $al$, J. Phys. G 37,075021(2010)

\bibitem{Lesi} T. Lesiak, hep-ex/0612042.

\bibitem{Rosn07}J. L. Rosner, J. Phys. G 34, S127 (2007) .

\bibitem{Paul}M. Paulini, arXiv:0906.0808.

\bibitem{Klem10}E. Klempt and J. M. Richard, Rev. Mod. Phys. 82, 1095 (2010)

\bibitem{Matt} M. Mattson et al, Phys. Rev. Lett. 89, 112001(2002)

\bibitem{Oche} A.Ocherashvili $et$ $al$, Phys. Lett. B 628, 18 (2005)

\bibitem{Faes} A. Faessler, Th. Gutsche, M. A. Ivanov, J. G. Korner, V. E. Lyubovitskij, D. Nicmorus,
and K. Pumsa-ard, Phys. Rev. D 73, 094013 (2006)

\bibitem{Pate}B. Patel, A. K. Rai, and P. C. Vinodkumar, J. Phys. G 35, 065001 (2008); J. Phys.
Conf. Ser. 110, 122010 (2008)

\bibitem{LiuX} Xiang Liu, Zhi-Gang Luo, Zhi-Feng Sun, Phys.Rev.Lett. 104,122001(2010);Jun He, Xiang Liu, Phys.Rev.D 82,114029 (2010); Xiang Liu, H.X.Chen, Y.R.Liu, A. Hosaka, S.L. Zhu, Phys. Rev. D 77,
014031(2008)

\bibitem{ChenHX1}Hua-Xing Chen, Wei Chen, Qiang Mao $et$ $al$, Phys. Rev. D
91, 054034 (2015); Hua-Xing Chen, Qiang Mao,Atsushi Hosaka, $et$
$al$., Phys. Rev. D 94, 114016 (2016)

\bibitem{Chun}Chun Mu, Xiao-Wang, Xiao-Lin Chen $et$ $al$, Chin. Phys. C 38, 113101
(2014)

\bibitem{ChenW}Wei Chen, Shi-Lin Zhu, Phys.Rev.D 81, 105018(2010); Wei Chen, Hong-ying Jin, R.T. Kleiv, $et$ $al$, Phys.Rev. D 88, 045027(2013)

\bibitem{Zhjr}J.R. Zhang, M.Q.Huang, Phys.Rev.D78, 094015(2008); Phys. Lett. B 674, 28(2009);  Chin. Phys. C 33, 1385(2009)

\bibitem{Karl1}M.Karliner, H.J.Lipkin,Phys. Lett. B 660, 539(2008)

\bibitem{Karl2}M.Karliner, J.L.Rosner,Phys. Rev. D 90, 094007(2014)

\bibitem{Nava} F.S.Navarra, M.Nielsen, K.Tsushima,Phys. Lett. B 606, 335(2005)

\bibitem{Khod} A.Khodjamirian, Th.Mannel, N.Offen, Y.M.Wang, Phys. Rev.
D83, 094031(2011)

\bibitem{Alie}T.M.Aliev, K.Azizi, M.Savc{\i},J. Phys. G 40, 065003(2013); 41,
065003(2014); J. Phys.: Conf. Ser. 556, 012016 (2014)

\bibitem{Azi1}K.Azizi, Y. Sarac, H.Sundu, Phys. Rev. D 90, 114011(2014)

\bibitem{Azi2}K.Azizi, Y.Sarac, H.Sundu, Nucl. Phys. A 943, 159(2015)

\bibitem{Wzg}Z.G.Wang,Phys. Rev. C 85,0 45204(2012) ;Eur.Phys.J.C 71, 1816(2011);Eur.Phys.J.C 73, 2533(2013)

\bibitem{KangXW}Xian-Wei Kang, J.A.Oller, arXiv:1612.08420[hep-ph]

\bibitem{Esposito}A. Esposito, A. Pilloni, and A.D. Polosa, Phys. Rept. 668, 1(2017)[arXiv:1611.07920]


\bibitem{Wzg5}Z.G. Wang, J.F. Li, Eur.Phys.J.A 47, 23(2011);Z.G.Wang, Phys.Rev.D 89, 034017(2014)

\bibitem{Navar}F. S. Navarra, M. Nielsen, Phys. Lett. B 443, 285(1998)

\bibitem{Khodj} A. Khodjamirian, Ch. Klein, and Th. Mannel, $et$ $al$, J. High Energy Phys.
09, 106(2011)

\bibitem{GuoLY2}G.L.Yu, Z.G.Wang, Z.Y.Li, arXiv:1608.03460[hep-ph]

\bibitem{Azizi4}K.Azizi, Y.Sarac and H.Sundu, Eur. Phys. J. A 52,4:114(2016)[arXiv:1510.05432]

\bibitem{Brac1}M.E. Bracco, M. Chiapparini, F.S. Navarra, M. Nielsen, Phys. Lett.
B 659, 559(2008)

\bibitem{Brac2}M.E. Bracco, M. Nielsen, Phys. Rev. D 82, 034012(2010)

\bibitem{Alie2}T.M. Aliev, M. Savc{\i}, arXiv:1308.3142 [hep-ph]

\bibitem{Alie3}T.M. Aliev, M. Savc{\i}, arXiv:1409.5250 [hep-ph]

\bibitem{Alie4}T.M. Aliev, T. Barakat, M. Savc{\i}, Phys. Rev. C 95, 035210(2017)

\bibitem{Doi}T. Doi, Y. Kondo, M. Oka, Phys. Rep. 398, 253(2004)

\bibitem{Altm}R. Altmeyer, M. Goeckeler, R. Horsley $et$ $al$, Nucl. Phys. Proc.
Suppl. 34, 373 (1994)

\bibitem{Wzg3}Z.G. Wang, S.L. Wan, Phys. Rev. D 74, 014017(2006)

\bibitem{Cerq}A. Cerqueira Jr, B.O. Rodrigues, M.E. Bracco, Nucl. Phys. A 874,
130(2012)

\bibitem{Rodr} B.O. Rodriguesa, M.E. Braccob, M. Chiapparinia, Nucl. Phys. A
929, 143(2014)

\bibitem{Yazi}E. Yazici $et$ $al$, Eur. Phys. J. Plus. 128(10), 113(2013)

\bibitem{Khos1}R. Khosravi, M. Janbazi, Phys. Rev. D 87, 016003(2013)

\bibitem{Khos2}R. Khosravi, M. Janbazi, Phys. Rev. D 89, 016001(2014)

\bibitem{Pasc}P. Pascual, R. Tarrach, Lect. Notes Phys. 194, 1(1984)

\bibitem{Wzg4}Z.G. Wang, Z.Y. Di, Eur. Phys. J. A 50, 143(2014)

\bibitem{Rein}L.J. Reinders, H. Rubinstein, S. Yazaki, Phys. Rep. 127, 1
(1985)

\bibitem{Guoly}Guo-Liang Yu, Zhen Yu Li, Zhi-Gang Wang, Eur. Phys. J. C 75, 243 (2015)

\bibitem{WzgH}Z.G.Wang, T.Huang,Phys. Rev. C 84, 048201(2011)

\bibitem{Kuma}A. Kumar, Adv. High Energy Phys., 549726(2014)

\bibitem{Haya}A. Hayashigaki, Phys. Lett. B 487, 96(2000)

\bibitem{Ioff1}B.L.Ioffe, Nucl.Phys.B 188, 317(1981)

\bibitem{Cola}P.Colangelo, A. Khodjamirian, At the frontier of particle
physics. in Handbook of QCD, vol. 3. (World Scientific, Singapore,
2000), p. 1495. arXiv:hep-ph/0010175)

\bibitem{WangZG5}Z.G.Wang, Eur.Phys.J.C 57, 711(2008); Eur.Phys.J.C
61,299(2009)

\bibitem{Oliv}K. A. Olive $et$ $al$,(Particle Data Group), Chin. Phys. C 38, 090001(2014)

\bibitem{Ioff2}B. L. Ioffe, Prog. Part. Nucl. Phys. 56, 232(2006)

\bibitem{Bely}V. M. Belyaev, B. L. Ioffe, Sov. Phys. JETP 57, 716(1983);
Phys. Lett. B 287, 176(1992)

\bibitem{WangZG}Z.G.Wang, Phys. Rev. C 85, 045204(2012)

\bibitem{Azi3}K. Azizi, N. Er, Eur. Phys. J. C 74, 2904(2014)

\bibitem{Khod4}A. Khodjamirian, ¡°B and D Meson Decay Constant in QCD,¡± in Proceeding of 3rd
Belle Analysis School, 22, 2010 (KEK, Tsukuba, Japan, 2010)

\bibitem{Eise}B. I. Eisenstein $et$ $al$, (CLEO Collab.), Phys. Rev. D 78,
052003(2008)
\end{thebibliography}
\end{document}